\documentclass[aps,pre,amsmath,amssymb,twocolumn,floatfix]{revtex4}

\usepackage{graphicx}
\newcommand{\R}{\mathbb{R}}
\newcommand{\iu}{\mathrm{i}} % imaginary unit
\begin{document}

\title{Driven transparent quantum graphs}
\author{J.R.~Yusupov$^{1}$, M.~Ehrhardt$^{2}$, Kh.Sh.~Matyokubov$^{3}$ and D.U.~Matrasulov$^{4}$}
\affiliation{$^1$Kimyo International University in Tashkent, 156 Usman Nasyr Str., 100121, Tashkent, Uzbekistan\\
$^2$Bergische Universit\"at Wuppertal, Gau{\ss}strasse 20, D-42119 Wuppertal, Germany\\
$^3$Urgench State Pedagogical Institute, 1 Gurlan str., 220100, Urgench, Uzbekistan\\
$^4$Turin Polytechnic University in Tashkent, 17 Niyazov Str., 100095, Tashkent, Uzbekistan}

%%%%%%%%%%%%%%%%%%%%%%%%%%%% Abstract
\begin{abstract}
% We consider quantum graphs interacting with an external potential. 
%By imposing transparent boundary conditions to the Schr\"odinger equation on a graph in the presence of an external field potential, we consider the reflectionless transmission of particles through the graph vertex. 
%
In this paper, we discuss the concept of quantum graphs with transparent vertices by considering the case where the graph interacts with an external time-independent field. 
In particular, we address the problem of transparent boundary conditions for quantum graphs, building on previous work on transparent boundary conditions for the stationary Schr\"odinger equation on a line. 
Physically relevant constraints making the vertex transparent under boundary conditions in the form of (weight) continuity and Kirchhoff rules are derived using two methods, the scattering approach and transparent boundary conditions for the time-independent Schr\"odinger equation. 
The latter is derived by extending the transparent boundary condition concept to the time-independent Schr\"odinger equation on driven quantum graphs.
We also discuss how the eigenvalues and eigenfunctions of a quantum graph are influenced not only by its topology, but also by the shape/type of a potential when an external field is involved.
\end{abstract}

\maketitle

%%%%%%%%%%%%%%%%%%%%%%%
\section{Introduction}
The concept of quantum graphs appeared in the lite\-rature about two decades ago. 
The first studies of branched quantum wires described using the Schr\"odinger equation on metric graphs came from Exner, Seba and Stovicek \cite{Exner1}. 
Later, Kostrykin and Schrader \cite{Kost} developed a more rigorous mathematical study of the problem by deriving general self-adjoint vertex boundary conditions. 
Various aspects of quantum graph theory have been considered in physical \cite{Uzy1,Uzy2,Gaspard,Uzy3,Uzy4,Alex,Bolte,Harrison,PTS,Keating} and mathematical contexts \cite{Kuchment04,Grisha1,Grisha,Kotlyarov,Mugnolo,Exner15}. Experimental realization of quantum graph spectra in microwave networks was presented in \cite{Hul,Lawn,dietz,drinko,Xiao}.
Such experimental study is possible because such networks can ``mimic'' quantum graphs, since waves there are described by the Helmholtz equation. Quantum graphs can be defined as a system of quantum wires connected according to a certain rule, called the \textit{topology of a graph}. 

Particles and waves in quantum graphs can be described by the stationary or time-dependent Schr\"odinger equation on metric graphs.
In such an approach, the wave function is a multi-component vector, with each component associated with a particular bond, and the Schr\"odinger equation is defined on each bond of a graph. 
To solve this equation, boundary conditions (matching conditions) must hold at the nodes of the graph, i.e., at the vertices, as well as at the end of each bond. 

In the case of the stationary Schr\"odinger equation, the vertex boundary conditions are used to obtain a secular equation for determining the eigenvalues of the problem. 
Depending on the topology of a graph, one obtains different eigenvalues, i.e., the eigenvalues and eigenfunctions of a quantum graph depend on its topology. 
When a quantum graph is subjected to the interaction of an external field, the solution of the problem depends not only on the topology but also on the shape/type of a potential.

In this paper we address the problem of \textit{transparent boundary conditions} for quantum graphs by considering these latter as interacting with an external time-independent potential, i.e.\ driven quantum graphs.
To formulate and solve the problem, we use the concept of transparent boundary conditions for the time-independent Schr\"odinger equation, which was previously considered in \cite{Klein2011,KleinThesis,Antoine10,review,moschini,Moyer2006,Klein2010}. 
In \cite{Klein2011} the extension of absorbing boundary conditions in the time domain to the computation of steady states is studied. 
The concept of transparent boundary conditions for the time-independent Schr\"odinger equation implies that its solutions (in the presence of an external time-independent potential) in the finite interval are equivalent to those in the whole space. 

We note that driven quantum graphs are less studied topic in the context of quantum graph theory, despite its importance for modeling quantum transport in interacting low-dimensional branched systems. 
In \cite{Uzy16} the trace formula for driven quantum graphs has been studied. In \cite{Bolte2,Bolte3} two-particle quantum graphs, interacting via the delta-potential is studied.
The ref.~\cite{Kurasov} considers optimal potentials for quantum graphs. 
Previously, quantum graphs driven by the Coulomb potential were used to model exciton transport in branched conducting polymers \cite{Hikmat20}. 
Quantum graphs confined in the harmonic oscillator potential were considered in \cite{Jambul1}.

This paper is organized as follows. In Section~\ref{sec::TBC1D}, we recall the problem of transparent boundary conditions for the stationary Schr\"odinger equation on a line. 
Section~\ref{sec:3} presents an extension of the problem for graphs. 
Section~\ref{sec:4} presents numerical experiments. 
Finally, concluding remarks are made in Section~\ref{sec:5}.

%%%%%%%%%%%%%%%%%%%%%%%
\section{Transparent boundary conditions for the stationary Schr\"{o}dinger equation on a line}\label{sec::TBC1D}
Here, following the Ref.~\cite{Klein2011}, we briefly recall the derivation of the transparent boundary conditions for the stationary Schr\"odinger equation on a real line. 
Consider the one-dimensional \textit{stationary Schr\"{o}dinger equation}
\begin{equation}\label{eq::1}  
    \Bigl(-\frac{\mathrm{d}^2}{\mathrm{d}x^2}+V\Bigr)\varphi
    =E\varphi, \quad x\in \R,  
\end{equation}
with a given, (generally nonlinear) potential $V:=V(x,\varphi)$. 
The solution of this problem is a pair $(\varphi,E)$, namely the eigenvalues $E$ and the corresponding eigenfunctions $\varphi$ of the system. 
Let us now solve the Eq.~\eqref{eq::1} in an interval with some boundary conditions, 
so that its solution coincides with the solution of the problem for the whole space restricted to the considered interval. 
In the following, such boundary conditions are called \textit{transparent boundary conditions (TBCs)}.

For the derivation of TBCs (also called absorbing boundary conditions in the literature) for stationary Schr\"{o}dinger equations, we refer to the work \cite{Klein2011}. 
Here we first consider the time-dependent problem on the whole space given as
\begin{equation}\label{eq::2}
    \begin{cases}
    \iu\partial_tu+\partial_x^2u+Vu=0, & \forall(x,t)\in \R\times\R^+, \\
    u(x,0)=u_0(x), & x\in\R.
    \end{cases}
\end{equation}
We also consider a finite computational interval $\Omega=[x_l,x_r]$. 
The artificial boundary $\Sigma$ is given by the two interval endpoints $x_l$ and $x_r$ and the outward unit normal vector to $\Omega$ is denoted by ${\bf n}$. 
% The outwardly directed unit normal vector to the bounded computational domain $\Omega=[x_l;x_r]$ is denoted by ${\bf n}$.
% The outwardly directed unit normal vector to the domain $\Omega$ is denoted by ${\bf n}$.

For this case the following second-order \eqref{eq::3} and fourth-order \eqref{eq::4} TBCs on the boundary $\Sigma$ were obtained:
\begin{equation}\label{eq::3}
    \partial_{\bf n}u = \iu\sqrt{\iu\partial_t+V} u, 
\end{equation}
\begin{equation}\label{eq::4}
   \partial_{\bf n}u = \iu\sqrt{\iu\partial_t+V}u
    + \frac{1}{4}\frac{\partial_{\bf n}V}{\iu\partial_t+V}u,
\end{equation}
where we take the square root with a positive real part.
In case of an operator $A$, $\sqrt{A}$ denotes the square root operator of $A$ with respect to its spectral decomposition. 

Now, using the transformation (`gauge transform') \cite{review}
\begin{equation}\label{eq::th}
    u(x,t)=\varphi(x)\,e^{-\iu Et},
\end{equation}
we obtain the time derivative
\begin{equation*}
   \iu\partial_t u=E\varphi(x)\,e^{-\iu Et},     
\end{equation*}
and thus the left hand side of \eqref{eq::3} reads
\begin{equation*}
   \iu\sqrt{\iu\partial_t+V}\, u = \sqrt{E-V}\,\bigl(\varphi(x)\,e^{-\iu Et}\bigr).  
\end{equation*}%
These considerations lead to some stationary TBCs corresponding to the Eqs.~\eqref{eq::3} and \eqref{eq::4}:
\begin{equation}
     \partial_{\bf n}\varphi = \iu \sqrt{E-V}\, \varphi, \quad\text{on} \,\,\Sigma,\label{tbc1d1}
\end{equation}
\begin{equation}\label{tbc1d2}
   \partial_{\bf n}\varphi = \iu\sqrt{E-V}\, \varphi
    + \frac{1}{4}\frac{\partial_{\bf n}V}{E-V}\, \varphi, \quad\text{on}\,\,\Sigma.
\end{equation}
In this way, the stationary TBCs
% transparent boundary conditions 
were derived assuming that the solution of \eqref{eq::2} can be written as a time-harmonic solution \eqref{eq::th}.

%%%%%%%%%%%%%%%%%%%%%%%
\subsection{Example: Confined quantum harmonic oscillator}
Here we demonstrate practical applications of TBC for realistic physical systems. 
Namely, we consider a quantum particle confined in a 1D box $[x_l,x_r]$ subjected to the interaction of a harmonic oscillator (parabolic) potential. 
Such a potential represents a parabolic well (caused by, e.g.\ external constant uniform magnetic field) with the cut-off at the box walls. 
The Hamiltonian of the system can be written as (in the units $\hbar=m=1$):
\begin{equation}
    H=-\frac{1}{2}\frac{\mathrm{d}^2}{\mathrm{d}x^2}+\frac{1}{2}\omega^2x^2.
\end{equation}

The complete solution of the stationary Schr\"odinger equation on the real line
\begin{equation}\label{eq:SE_stat}
   H\phi=E\phi,\quad x\in\R,
\end{equation}
can be written in terms of the confluent hypergeometric function of the first kind:
\begin{multline}
  \phi(x)=\exp\Bigl(-\frac{1}{2} \omega x^2\Bigr)
    \biggl[ A\,M\Bigl(\frac{1}{4}-\frac{E}{2\omega};\frac{1}{2};\omega x^2\Bigr)\\ 
     + B\,\omega^\frac{1}{2} x
     M\Bigl(\frac{3}{4}-\frac{E}{2\omega};\frac{3}{2};\omega x^2\Bigr) \biggr]
\end{multline}
where $A$ and $B$ are constants and $M(\cdot;\cdot;\cdot)$ denotes the \textit{Kummer's function} \cite{Abramowitz}. 
For asymptotic boundary conditions imposed as $\phi(|x|\to\infty) =0$, the solutions are the Hermite functions and the corresponding eigenvalues can be written as 
\begin{equation}
     E_n = \omega \Bigl(n + \frac{1}{2}\Bigr).
\label{eigen}
\end{equation}

Imposing the TBCs \eqref{tbc1d1} results in a system of linear algebraic equations with respect to the coefficients $A$ and $B$:
\begin{equation}\label{eq::syst}
\mathbf{h}(E)\begin{pmatrix}A\\B\end{pmatrix}=0,
\end{equation}
where the elements of the $2\times2 $ matrix $\mathbf{h}$ are given as
\begin{multline*}
    h_{11}=-\Bigl(\omega x_l+\iu\sqrt{E-\frac{1}{2}\omega^2x_l^2}\Bigr)  M\Bigl(\frac{1}{4}-\frac{E}{2\omega};\frac{1}{2};\omega x_l^2\Bigr)\\
    +\Bigl(\frac{1}{2}-\frac{E}{\omega}\Bigr) M\Bigl(\frac{5}{4}-\frac{E}{2\omega};\frac{3}{2};\omega x_l^2\Bigr)
\end{multline*}

\begin{multline*}
   h_{12}=-\Bigl(\omega x_l^2-1+\iu x_l\sqrt{E-\frac{1}{2}\omega^2x_l^2}\Bigr)  M\Bigl(\frac{3}{4}-\frac{E}{2\omega};\frac{3}{2};\omega x_l^2\Bigr)\\
    +x_l^2\Bigl(\frac{1}{2}-\frac{E}{3\omega}\Bigr) M\Bigl(\frac{7}{4}-\frac{E}{2\omega};\frac{5}{2};\omega x_l^2\Bigr)
\end{multline*}

\begin{multline*}
h_{21}=-\Bigl(\omega x_r-\iu\sqrt{E-\frac{1}{2}\omega^2x_r^2}\Bigr)  M\Bigl(\frac{1}{4}-\frac{E}{2\omega};\frac{1}{2};\omega x_r^2\Bigr)\\
+\Bigl(\frac{1}{2}-\frac{E}{\omega}\Bigr) M\Bigl(\frac{5}{4}-\frac{E}{2\omega};\frac{3}{2};\omega x_r^2\Bigr)
\end{multline*}

\begin{multline*}
   h_{22}=-\Bigl(\omega x_r^2-1+\iu x_r\sqrt{E-\frac{1}{2}\omega^2x_r^2}\Bigr)  M\Bigl(\frac{3}{4}-\frac{E}{2\omega};\frac{3}{2};\omega x_r^2\Bigr)\\
   +x_r^2\Bigl(\frac{1}{2}-\frac{E}{3\omega}\Bigr)  M\Bigl(\frac{7}{4}-\frac{E}{2\omega};\frac{5}{2};\omega x_r^2\Bigr)
\end{multline*}

The existence of non-trivial solutions of the above algebraic system \eqref{eq::syst} leads to the following secular equation with respect to $E$
\begin{equation}\label{eq::syst2} 
    \det\bigl(\mathbf{h}(E)\bigr)=0.
\end{equation}
The roots of Eq.~\eqref{eq::syst2} give the eigenvalues of the confined quantum harmonic oscillator with TBCs.
These eigenvalues and corresponding eigenvectors must coincide with the solution of the whole-space problem restricted to $[x_l,x_r]$. 
To show this, the Table~\ref{tab::ener} shows the first five numerically computed eigenvalues (energies) of the considered system for $\omega=1$ and interval $[-5,5]$. 
 
\begin{table}[h]
\caption{The first five eigenvalues of the confined quantum harmonic oscillator (to be compared with those of unconfined one given by Eq.~\eqref{eigen}) for which the transparent boundary conditions \eqref{tbc1d1} are imposed at $x_l=-5$ and $x_r=5$ ($\omega=1$).}\label{tab::ener}
\begin{tabular}{|c|l|}
\hline
\multicolumn{1}{|c|}{\ $n$ \ } & \multicolumn{1}{c|}{$E_n$} \\ \hline
1                       & 0.50000000005979         \\ \hline
2                       & 1.49999999734821         \\ \hline
3                       & 2.50000006391437         \\ \hline
4                       & 3.49999862590053          \\ \hline
5                       & 4.50000910443786          \\ \hline
\end{tabular}
\end{table}

We note that more detailed analysis of this example is done in the Ref.~\cite{Klein2011} by performing a numerical test which consists in presenting the error on both the energy and eigenfunctions depending on the computational domain size. 
Furthermore, another case of the P\"oschl-Teller potential is also considered in that work. 
In \cite{KleinThesis,Antoine10}, the case of the Morse potential as well as the Woods-Saxon potential is also studied. 
These examples extend the conclusions on TBCs \eqref{tbc1d1} and \eqref{tbc1d2}.

%%%%%%%%%%%%%%%%%%%%%%%%%%%%%%%%%%%%%%%%%%%
\section{Quantum graphs with transparent vertices}\label{sec:3}

%%%%%%%%%%%%%%%%%%%%%%%%%%
\subsection{The scattering approach}
The concept of transparent boundary conditions has recently been extended to the case of evolution equations on graphs, by considering linear  \cite{Jambul} and nonlinear \cite{Jambul2} Schr\"odinger equations on graphs and the Dirac equation on quantum graphs \cite{Jambul20}.
In all cases the focus was on time-dependent equations.

Here we consider the Schr\"odinger equation on the star graph given on each of its bonds as
\begin{equation}\label{se}
    -\frac{\mathrm{d}^2}{\mathrm{d}x^2}\Psi_j(x)
    =k^2\Psi_j(x),\quad j=1,2,\dots,N,
\end{equation}
where $N$ is the number of bonds.
The coordinates $x$ take the value $0$ at the vertex and go to infinity at the unbounded ends.

The wave function must satisfy certain boundary conditions at the vertices, the imposition of which 
guarantees that the resulting Schr\"odinger operator is self-adjoint. 
One of these boundary conditions at the vertex can be written as a generalized form of the continuity condition
\begin{equation}\label{bc1}
    \alpha_1\Psi_1(0)=\alpha_2\Psi_2(0)=\ldots=\alpha_N\Psi_N(0),
\end{equation}
and the current conservation
\begin{equation}\label{bc2}
     \sum\limits_{j=1}^N{\frac{1}{\alpha_j}\frac{\mathrm{d}}{\mathrm{d}x}\Psi_j(0)=0},
\end{equation}
where $\alpha_j\in\R$ are some non-zero coefficients.

Using the scattering approach \cite{Uzy1}, the total wave function ${\bf \Psi}$ can be written as a linear combination of functions $\Psi_i$,
which are solutions for the case where an incoming wave enters the vertex from bond $i$ 
and outgoing waves from the vertex to all bonds $j$ (including $j=i$ corresponding to the reflected part). 
$\Psi_i$ is an $N$-dimensional vector with components $\Psi_{i,j}(x)$ for all $1\le j\le N$,
\begin{equation}\label{wf}
    \Psi_{i,j}(x)=\delta_{j,i}\,e^{-\iu kx} + \sigma_{i,j}\,e^{\iu kx}.
\end{equation}
Here $\sigma$ is the $N\times N$ scattering matrix, which provides a transformation between the incoming and the outgoing waves at the vertex. 
The continuity condition \eqref{bc1} together with \eqref{wf} results in
\begin{equation}\label{rbc1}
    \sigma_{i,j}=\frac{\alpha_i}{\alpha_j}(1+\sigma_{i,i}), 
    \quad\forall j\,\,(j\neq i).
\end{equation}
Now, the current conservation condition \eqref{bc2} together with \eqref{wf} leads to
\begin{equation}
   1-\sigma_{i,i}=\sum_{\footnotesize \begin{array}{c}
                              j=1 \\
                              j\neq i
                            \end{array}}^N{\frac{\alpha_i}{\alpha_j}\,\sigma_{i,j}}.
\end{equation}
Then, using \eqref{rbc1} we can determine $\sigma_{i,i}$
%\begin{equation}
%\sigma_{i,i}=\frac{1-\sum_{\tiny \begin{array}{c}
%                              j=1 \\
%                              j\neq i
%                            \end{array}}^N{\frac{\alpha_i^2}{\alpha_j^2}}}{1+\sum_{\tiny \begin{array}{c}
%                              j=1 \\
%                              j\neq i
%                            \end{array}}^N{\frac{\alpha_i^2}{\alpha_j^2}}}.\label{sigma}
%\end{equation}
\begin{equation}\label{sigma}
\sigma_{i,i}=
\Bigl(1-\sum\limits_{\tiny \begin{array}{c}
                              j=1 \\
                              j\neq i
                            \end{array}}^N{\frac{\alpha_i^2}{\alpha_j^2}}\Bigr)/
                            \Bigl(1+\sum\limits_{\tiny \begin{array}{c}
                              j=1 \\
                              j\neq i
                            \end{array}}^N{\frac{\alpha_i^2}{\alpha_j^2}}\Bigr).
\end{equation}

Since we are interested in non-reflecting waves, we need to make the reflection probability $|\sigma_{i,i}|^2$ vanish. 
Obviously, this can be achieved for the \eqref{sigma} by satisfying the following conditions on the coefficients
$\alpha_j$, $(j=1,2,\dots,N)$, in the form of a sum rule
\begin{equation}\label{sumrule}
\frac{1}{\alpha_i^2}=\sum_{\tiny \begin{array}{c}
                              j=1 \\
                              j\neq i
                            \end{array}}^N{\frac{1}{\alpha_j^2}}.
\end{equation}
The satisfaction of \eqref{sumrule} corresponds to the case where an incoming wave from bond $i$ enters the vertex without reflection and outgoing waves from the vertex to all other bonds $j$. 
Note that this condition \eqref{sumrule} was derived earlier by applying the transparent boundary conditions to the time-dependent Schr\"odinger equation on graphs \cite{Jambul}.

Now, let us consider the same problem in presence of some potential $V_j(x)$ given for each bond $j$\,:
\begin{equation}\label{seV}
    -\frac{\mathrm{d}^2}{\mathrm{d}x^2}\Psi_j(x)+V_j(x)\Psi_j(x)=k^2\Psi_j(x),
    \quad j=1,2,\dots,N,
\end{equation}
with the same boundary conditions \eqref{bc1} and \eqref{bc2}.
In this case, the components $\Psi_{i,j}(x)$ of the total wave function can be written as
\begin{equation}\label{wfV}
     \Psi_{i,j}(x)=\delta_{j,i}\,e^{-\iu\sqrt{k^2-V_j(x)}x} 
                  +\sigma_{i,j}\,e^{\iu\sqrt{k^2-V_j(x)}x}.
\end{equation}

The boundary conditions \eqref{bc1} and \eqref{bc2} at the vertex together with \eqref{wfV} can be used to determine $\sigma_{i,i}$
\begin{equation}\label{sigmaV}
\sigma_{i,i}=\frac{1-\sum_{\tiny \begin{array}{c}
                              j=1 \\
                              j\neq i
                            \end{array}}^N{\frac{\alpha_i^2}{\alpha_j^2}\sqrt{\frac{k^2-V_j(0)}{k^2-V_i(0)}}}}{1+\sum_{\tiny \begin{array}{c}
                              j=1 \\
                              j\neq i
                            \end{array}}^N{\frac{\alpha_i^2}{\alpha_j^2}\sqrt{\frac{k^2-V_j(0)}{k^2-V_i(0)}}}}.
\end{equation}
Again, by requiring the reflection probability to vanish %be equal to zero 
we get the energy dependent expression
\begin{equation}\label{sumruleV}
\frac{\sqrt{k^2-V_i(0)}}{\alpha_i^2}=\sum_{\tiny \begin{array}{c}
                              j=1 \\
                              j\neq i
                            \end{array}}^N
                            \frac{\sqrt{k^2-V_j(0)}}{\alpha_j^2}.
\end{equation}
Under the condition that all the bond potentials have the same value at the vertex (i.e.\ $V_1(0)=V_2(0)=\dots=V_N(0)$), we can obtain the same constraint given by \eqref{sumrule}.

%%%%%%%%%%%%%%%%%%%%%%%%%%
\subsection{Derivation of the sum rule using the transparent boundary conditions}

Here we use the concept of transparent boundary conditions  (TBCs) introduced in Section~\ref{sec::TBC1D}, to derive the conditions for the vertex transparency. 
In particular, we focus on the solution of Eq.~\eqref{seV} with boundary conditions \eqref{bc1} and \eqref{bc2} imposed on the vertex.

Assuming that the solution of the time-dependent version of Eq.~\eqref{seV} can be written as a time-harmonic solution
\begin{equation}\label{eq::thG}
    \Psi_j(x,t)=\varphi_j(x)\,e^{-\iu Et}, \quad j=1,2,\ldots,N,
\end{equation}
we can write the stationary transparent boundary conditions for all the bonds as
\begin{equation}\label{tbcG1}
     \partial_{\bf n}\varphi_j = \iu \sqrt{E-V_j}\, \varphi_j,
\end{equation}
for the second-order approximation and
\begin{equation}\label{tbcG2}
    \partial_{\bf n}\varphi_j 
    = \iu\sqrt{E-V_j}\, \varphi_j
     + \frac{1}{4}\frac{\partial_{\bf n}V_j}{E-V_j} \varphi_j, 
\end{equation}
for the fourth-order approximation at the vertex ($x=0$).

We now assume that the $i$th bond number is the computational domain and the other bonds are external bonds. 
Considering the higher order of the given approximations, i.e., the Eq.~\eqref{tbcG2} and using the current conservation conditions \eqref{bc2}, we obtain
\begin{equation}\label{cc}
   \frac{1}{\alpha_i}\frac{\mathrm{d}}{\mathrm{d}x}\varphi_i+ 
    \sum\limits_{\tiny \begin{array}{c}
                              j=1 \\j\neq i
                            \end{array}}^N{\frac{1}{\alpha_j}\Bigl(\iu\sqrt{E-V_j}\,\varphi_j
+ \frac{1}{4}\frac{V'_j}{E-V_j} \varphi_j\Bigr)}=0,
\end{equation}
where $V_j=V_j(0)$ and $V'_j=\frac{\mathrm{d}}{\mathrm{d}x}V_j(x)|_{x=0}$. 
Using the continuity condition \eqref{bc1}, we can write Eq.~\eqref{cc} in terms of $\varphi_i$ as
\begin{equation}\label{cc1}
   -\frac{1}{\alpha_i}\partial_{\bf n} \varphi_i + \sum\limits_{\tiny \begin{array}{c}
                              j=1 \\
                              j\neq i
                            \end{array}}^N{\frac{\alpha_i}{\alpha_j^2}\Bigl(\iu\sqrt{E-V_j}\,\varphi_i
+ \frac{1}{4}\frac{V'_j}{E-V_j} \varphi_i\Bigr)}=0.
\end{equation}
At this point we assume that (i) $V_j(0)=V_i(0)$ and (ii) $V'_j(0)=-V'_i(0)$ for all bonds.
This assumptions lead us to
\begin{equation}\label{cc2}
   \partial_{\bf n} \varphi_i
   =\Bigl(\iu\sqrt{E-V_i}\, \varphi_i
+ \frac{1}{4}\frac{\partial_{\bf n}V_i}{E-V_i} \varphi_i\Bigr)\sum\limits_{\tiny \begin{array}{c}
                              j=1 \\
                              j\neq i
                            \end{array}}^N{\frac{\alpha_i^2}{\alpha_j^2}}.
\end{equation}
It is easy to see from Eq. \eqref{cc2} that when the condition \eqref{sumrule} is fulfilled, i.e.\ when
\begin{equation}\label{smrule}
\sum\limits_{\tiny \begin{array}{c}
                              j=1 \\
                              j\neq i
                            \end{array}}^N{\frac{\alpha_i^2}{\alpha_j^2}}=1,
\end{equation}
Thus we obtain the TBC for the bond with number $i$. 
Note that assumption (i) on the bond potentials is sufficient for the second-order approximation to derive the sum rule, 
while the inclusion of the second term in the fourth-order approximation additionally requires assumption (ii).

%%%%%%%%%%%%%%%%%%%%%%%%%%%%%%%%%%%%%%
\begin{figure}[t!]
\includegraphics[width=70mm]{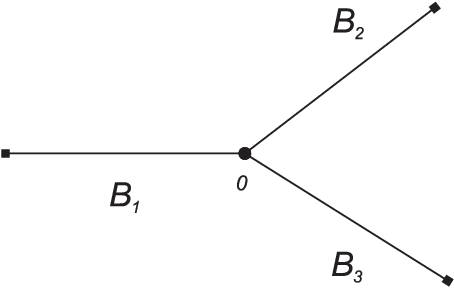}
\caption{Finite star graph with three bonds. 
The coordinates of the bonds $B_j$ take the value $0$ at the central vertex and $L_j$ at the edge vertices ($j=1,2,3$).}
\label{fig::star}
\end{figure}

Note that under the same assumptions (i) and (ii) above, the condition for transparent vertex boundary conditions \eqref{smrule} can be generalized for the case when there are $M<N$ incoming leads (e.g., bonds $j=1,2,\dots,M$) and $N-M$ outgoing leads (bonds $j=M+1,M+2,\dots,N$). For this case we can write
\begin{align}\label{cc3}
&\partial_{\bf n} \varphi_i
   =\Bigl(\iu\sqrt{E-V_i}\, \varphi_i
+ \frac{1}{4}\frac{\partial_{\bf n}V_i}{E-V_i} \varphi_i\Bigr)\cdot \nonumber\\ 
&\left(-\sum\limits_{\tiny \begin{array}{c}
                              j=1 \\
                              j\neq i
                            \end{array}}^M{\frac{\alpha_i^2}{\alpha_j^2}}+\sum\limits_{\tiny j=M+1}^N{\frac{\alpha_i^2}{\alpha_j^2}}\right),\text{ for } i=1,2,...,M.
\end{align}

Then, the general form of the constraint reads
\begin{equation}
    \sum\limits_{j=1}^M{\frac{1}{\alpha_j^2}}=\sum\limits_{j=M+1}^N{\frac{1}{\alpha_j^2}}.
\end{equation}

We note that extending the above treatment to the case of other graphs (e.g., tree, loop, triangle, etc.) can be done along the same lines as for the star graph.

%%%%%%%%%%%%%%%%%%%%%%%%%%%%%
\section{The spectrum of the driven quantum star graph with transparent vertex}\label{sec:4}

Consider the Schr\"odinger equation \eqref{se} on the star graph with the boundary conditions given by the equations \eqref{bc1} and \eqref{bc2} at the central vertex and the Dirichlet boundary conditions at the edge vertices. 
These boundary conditions ensure that the Schr\"odinger equation is self-adjoint and thus that an unbounded discrete spectrum exists. 
At each boundary $j$ ($j=1,2,\dots,N$), the wave function $\Psi_j$ can be written as 
\begin{equation}
    \Psi_j(x)=\frac{C_n}{\alpha_j \sin{kL_j}}\sin{k(L_j-x)},
\end{equation}
where the normalization coefficients are given by
\begin{equation}    
   C_n=\sqrt{2}\biggl(\sum_j{\frac{L_j+\sin{(2k_nL_j)}}{\alpha_j^2\sin^2{(k_nL_j)}}}\biggr)^{-\frac{1}{2}}
\end{equation}

%%%%%%%%%%%%%%%%%%%%%%%%%%%%%%%%%%%
\begin{figure}[t!]
\includegraphics[width=90mm]{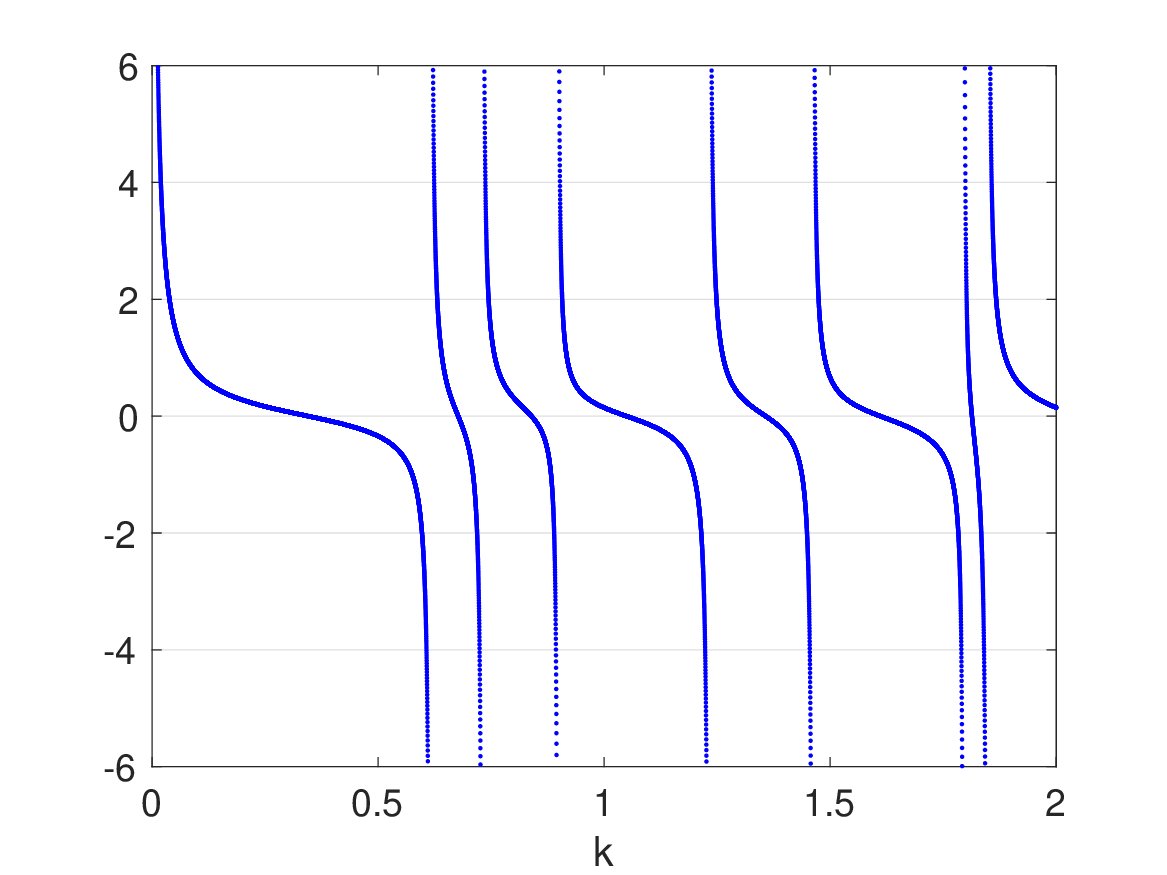}
\caption{The secular function \eqref{eq::sf} for the star graph is shown in Fig.~\ref{fig::star} with $L_1=5.1$, $L_2=4.3$, $L_3=3.5$ and boundary condition parameters $\alpha_1=2.4$, $\alpha_2=3$, $\alpha_3=4$.}
\label{fig::sf}
\end{figure}

The eigenvalues of the considered star graph are found by solving the secular equation
\begin{equation}\label{eq::sf}
    \sum\limits_{j=1}^{N}{\frac{1}{\alpha_j^2\tan{kL_j}}}=0.
\end{equation}

\begin{figure}[t!]
\includegraphics[width=90mm]{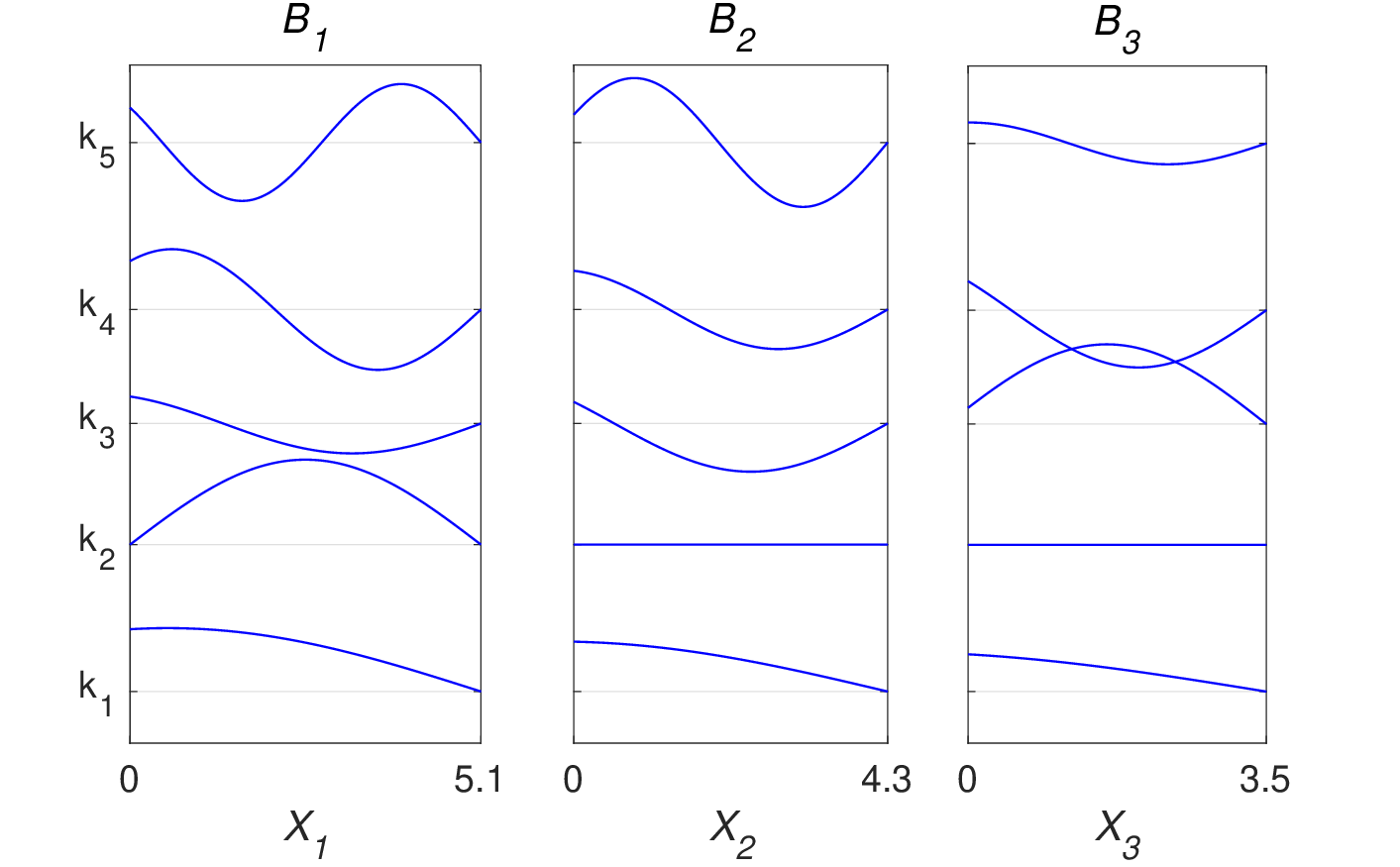}
\caption{Eigenfunctions corresponding to the first five eigenvalues.}
\label{fig::wf}
\end{figure}

Let us give an example. 
We consider the simple case of a star graph with three bonds of lengths $L_1=5.1$, $L_2=4.3$, $L_3=3.5$ (see Fig.~\ref{fig::star}).
We will compute the spectrum of the quantum star graph with boundary conditions \eqref{bc1} and \eqref{bc2} in the central vertex and Dirichlet boundary conditions in the edge vertices. 
To make the central vertex transparent, we choose parameters such as $\alpha_1=2.4$, $\alpha_2=3$, $\alpha_3=4$ that satisfy the sum rule \eqref{sumrule}. 
The eigenvalues of such a graph are found by computing the zeros of the secular function on the left side of Eq.~\eqref{eq::sf}. 
The graph of this function is shown in Fig.~\ref{fig::sf}. 
The first five eigenvalues are listed in Table~\ref{tab::eig}. 
The corresponding eigenfunctions are plotted in Fig.~\ref{fig::wf}.

\begin{table}[h]
\caption{Values of the first five eigenvalues of the considered quantum star graph.}\label{tab::eig}
\begin{tabular}{|c|l|}
\hline
\multicolumn{1}{|c|}{\ $n$ \ } & \multicolumn{1}{c|}{$k_n$} \\ \hline
1                       & 0.34519971576497         \\ \hline
2                       & 0.61599855952741         \\ \hline
3                       & 0.83908748094051         \\ \hline
4                       & 1.04930134453149          \\ \hline
5                       & 1.35717565248649          \\ \hline
\end{tabular}
\end{table}

%%%%%%%%%%%%%%%%%%%%%%%%%%%%%
\section{Conclusions}\label{sec:5}

We considered the problem of driven (interacting with external field) quantum graph with transparent vertex by combining the concept of transparent boundary conditions with the scattering approach. 
Transparent vertex here implies absence of backscattering at the vertex.
Transparent vertex boundary conditions for the time-independent Schr\"odinger equation on the graph containing external potential are derived. 
Using the scattering approach, it is shown that such conditions ensure the absence of backscattering at the vertex, provided that certain constraints on the system parameters are satisfied.

A more rigorous mathematical study of quantum graphs has been carried out considering vertex boundary conditions leading to a self-adjoint Schr\"odinger equation. 
Although the above results are obtained for the case of a star-branched graph, the extension of the approach to other graph topologies is rather trivial. 

The model of transparent quantum graph presented in this study can be directly applied to describe reflectionless quantum transport in various low-dimensional systems, such as branched molecular chains and conducting polymers, semiconductor quantum wire networks, multi-branched topological insulators, etc. 
The extension of the approach to other quantum mechanical wave equations, such as Dirac, Klein-Gordon and Bogoliubov de Gennes equations, is of importance and is the subject of future research.

% The eigenvalues and eigenfunctions of a quantum graph depend on its topology and on external fields.

 %%%%%%%%%%%%%%%%%%%%%%%%%%%%%%%%%%%
\begin{acknowledgments}
The work is supported by the grants of the Innovative development Agency under the Ministry of higher education, science and innovations of the Republic of Uzbekistan (Ref. No. F-2021-440 and FZ-5821512021).
\end{acknowledgments}

%\appendix

%\section{Appendixes}

%%%%%%%%%%%%%%%%%%%%%%%%%%%%%

%%%%%%%%%%%%%%%%%%%%%%%%%%%%%%%%%%%%%%%%%%%

\end{document}